\documentclass[aps,twocolumn,showpacs,preprintnumbers,amsmath,amssymb,nofootinbib,superscriptaddress,showkeys]{revtex4}

\usepackage{hyperref}
\usepackage{float}
\usepackage{epsfig}
\usepackage{graphicx}
\usepackage{mathptmx}      
\usepackage{latexsym}
\usepackage{natbib}
\usepackage{longtable}
\usepackage{dcolumn}

\DeclareMathVersion{nxbold}
\SetSymbolFont{operators}{nxbold}{OT1}{cmr} {b}{n}
\SetSymbolFont{letters}  {nxbold}{OML}{cmm} {b}{it}
\SetSymbolFont{symbols}  {nxbold}{OMS}{cmsy}{b}{n}

\DeclareMathAlphabet{\mathcal}{OMS}{cmsy}{m}{n}
\begin{document}

\title{Coarse graining of NN inelastic interactions up to
  3 GeV:\\ Repulsive vs Structural core}

\author{P. Fern\'andez-Soler} \email{pedro.fernandez@ific.uv.es}
\affiliation{Instituto de F\'{\i}sica Corpuscular (IFIC), Centro Mixto
  CSIC-Universidad de Valencia, \\ Institutos de Investigaci\'on de
  Paterna, Aptdo. 22085, 46071 Valencia, Spain} \author{E. Ruiz
  Arriola} \email{earriola@ugr.es} \affiliation{Departamento de
  F\'{\i}sica At\'omica, Molecular y Nuclear \\ and Instituto Carlos I
  de F{\'\i}sica Te\'orica y Computacional \\ Universidad de Granada,
  E-18071 Granada, Spain.}

\date{\today}

\begin{abstract} 
The repulsive short distance core is one of the main paradigms of
nuclear physics which even seems confirmed by QCD lattice
calculations. On the other hand nuclear potentials at short distances
are motivated by high energy behavior where inelasticities play an
important role. We analyze NN interactions up to 3 GeV in terms of
simple coarse grained complex and energy dependent interactions.  We
discuss two possible and conflicting scenarios which share the common
feature of a vanishing wave function at the core location in the
particular case of S-waves. We find that the optical potential with a
repulsive core exhibits a strong energy dependence whereas the optical
potential with the structural core is characterized by a rather adiabatic
energy dependence which allows to treat inelasticity perturbatively.
We discuss the possible implications for nuclear structure
calculations of both alternatives.
\end{abstract} 


\maketitle

\section{Introduction}

The nuclear hard core was postulated by Jastrow in
1951~\cite{Jastrow:1951vyc}, based on the observation that pp
scattering presents a flat and almost angle independent differential
cross section at about 100-200 MeV, a feature that he found to be
easily reproduced by a hard sphere of radius $a_c=0.6 {\rm fm}$.  This
finding has been corroborated by more complete analyses reaching
larger energies.  Actually, in his analysis above 1 GeV in 1958
G.E. Brown found that the core remained below this energy but made the
intriguing claim ``The hard core is assumed to disappear with
increasing energy and to be replaced by
absorption''~\cite{brown1958proton}. The repulsive core has become one
of the well accepted paradigms of nuclear physics providing a possible
explanation for nuclear stability for high density states. Also, the
short distance properties are relevant for neutron matter in the core
of a neutron star with a Fermi momentum about twice that of nuclear
matter, $ k_F \sim 600 {\rm MeV}$ and hence a corresponding reduced
wavelength $1/k_F \sim 0.3 {\rm fm}$. The nuclear physics
evidence~\cite{preston1975structure} for this prominent feature is
shown to be the zero crossing of the $^1S_0$ NN phase-shift at $T_{\rm
  LAB} \sim 300 {\rm MeV}$, right after the opening of the pion
production threshold. In line with these early developments, the
Hamada-Johnston potential became the archetype of a hard core
potential with a common core radius $a_c=0.5 {\rm
  fm}$~\cite{Hamada:1962nq} and the Reid potential confirmed these
findings with both a hard core and a soft core type
structure~\cite{Reid:1968sq,Tamagaki:1968zz} (see also
\cite{Otsuki:1969qu}) for a comprehensive review.

That was much of the discussion in those days in nuclear physics where
the core has a visible effect in the nuclear and neutron matter
equation of state~\cite{bethe2012nuclear}. A readable historical
account can be found in Ref.~\cite{Machleidt:1989tm}.  The usual
characterization of the hard core requires the use of local (or weakly
nonlocal) potentials. Indeed, the Argonne potential saga befits this
viewpoint~\cite{Lagaris:1981mm,Wiringa:1984tg,Wiringa:1994wb} and is
the natural evolution of these developments allowing benchmarking
calculations in nuclear structure of light nuclei (for reviews see
e.g.  Refs.~ \cite{Pieper:2001mp,Carlson:2014vla} and references
therein).

In this paper we want to analyze critically the evidences of the
nuclear core which ultimately proves crucial for nuclear structure and
nuclear reactions calculations at intermediate and high energies. We
do so by paying attention to the scattering process at those energies
probing the core size.

In order to motivate our analysis below it is pertinent to review
several aspects and features of the nuclear core within various
contexts. This is done in Section~\ref{sec:core} where we review some
of the history on the repulsive and structural cores and their
corresponding fingerprints. We also discuss critically aspects of the
problem based on recent lattice QCD results, as well as the more
phenomenological approaches which demand a realistic treatment of
relativity, inelasticity and spin degrees of freedom. An analysis of
the relevant scales in the problem is undertaken in
Section~\ref{sec:fluctuations} trying to be as pedagogical as
possible. There we show that the largest LAB energies where a partial
wave analysis (PWA) has been carried out in the past, actually probes
the region where the core sets in, but does not resolve the fine
structure of the core shape. In section \ref{sec:elastic} we present
some numerical results revisiting aspects of the coarse graining idea
in the elastic case for fits up to 300 MeV and 1 GeV.  Full
consideration of inelasticities is undertaken in
Section~\ref{sec:optical} where we extend the idea to the case of an
optical potential and provide fits up to 3GeV where we confront two
validated and conflicting scenarios: the repulsive core and the
structural core. In section \ref{sec:sc} we provide some discussion
and outlook for future work. Finally, in Section \ref{sec:concl} we
summarize our results and main conclusions.

\section{The Nuclear Short distance Core}
\label{sec:core}

\subsection{Early origins of Repulsive and structural core}

The origin of the repulsive core has been the subject of many
investigations. Within the One Boson Exchange (OBE) picture where
nucleons exchange all possible mesons, $\pi,\eta,\sigma,\rho,\omega,
\dots$ the repulsive core has traditionally been attributed to the
$\omega$-meson exchange after Nambu~\cite{Nambu:1957vw} and many
others~\cite{Bryan:1964zzb}, but with an unnatural coupling $g_{\omega
  NN} \sim 20$~\cite{Machleidt:1989tm} even in the extremely
successful CD-Bonn potential~\cite{Machleidt:2000ge} which largely
violates SU(3) expectations, $g_{\omega NN} = 3 g_{\rho NN} \sim 10$,
and would represent a unrealistically large deviation (about $50\%$)
from this symmetry~\footnote{For the role played by $2\pi$ exchange as
  a scalar meson see e.g. Ref.~\cite{Oset:2000gn} and references
  therein .}. A relativistic origin of the core was prompted by
Gross~\cite{Gross:1972ye}.

An alternative origin of the repulsive core, the so-called structural
core~\cite{Otsuki:1965yk,TAMAGAKI:1967zz,Otsuki:1969qu} was proposed
many years ago and is based on the composite nature of the nucleon and
the Pauli principle at the constituent level which implies that the
zero energy wave function has a zero at the core radius but does not
vanish below it. This implies the existence of forbidden deeply bound
states ~\cite{Neudatchin:1975zz}, a fact accommodated naturally by
quark cluster models~\cite{Neudachin:1977vt} that has motivated the
series of Moscow potentials~\cite{Kukulin:1992vd}. The connection with
forbidden states on the light of high energy scattering data below 6
GeV was addressed in Ref.~\cite{Neudachin:1991qn}. A readable and
fresh account on these well documented Short-range components of
nuclear forces can be found in a recent paper by Kukulin and
Platonova~\cite{Kukulin:2013oya}.  It is noteworthy that within the
OBE picture it has been found that if a SU(3)-natural coupling
$g_{\omega NN} = 3 g_{\rho NN} \sim 10$ is assumed the
$\omega$-exchange repulsion is overcome by attractive $\rho$ and
$\sigma$-exchanges triggering a net short distance strong attraction
and a spurious deeply bound state in the $^1S_0$ channel is
generated. As a consequence of the oscillation theorem the
corresponding wave function at zero energy develops a node which is
located at about the standard and traditionally accepted core
position~\cite{Cordon:2009pj}.

\subsection{The nuclear core from QCD }

The ultimate answer on the existence of the core and its properties
should come from QCD. In fact, recent QCD Lattice calculations claim
to find a repulsive core at about similar distances in the quenched
approximation~\cite{Aoki:2011ep,Aoki:2013tba}. This is done by placing
two heavy sources made of three quark fields with nucleon quantum
numbers at the same point, $J_N(x) \sim q(x) q(x) q(x)$, located
at a given fixed separation $\vec r$ and studying the propagation of
the corresponding correlators for long enough Euclidean times
providing the corresponding static energy $E_{NN}(r) \sim 2 M_N +
V_{NN} (r)$. Moreover, an application of the Operator Product
Expansion provides understanding of short distance $V_{NN} (r) \sim
1/( M_N r^2)$-type repulsions among point-like baryons in
QCD~\cite{Aoki:2012xa}.

\begin{figure}[t]
\begin{center}
\epsfig{figure=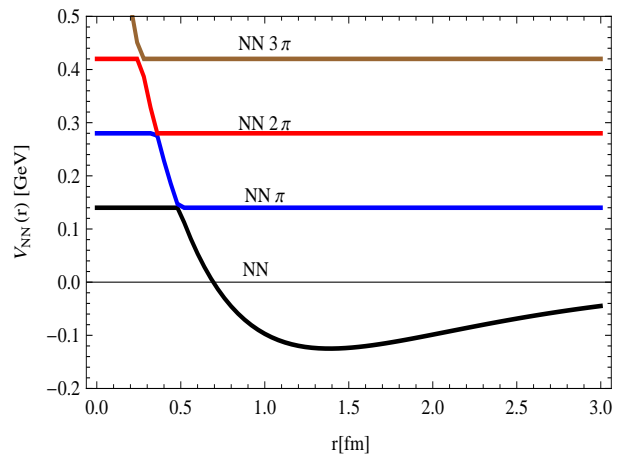,width=8cm,height=6cm} 
\end{center}
\caption{(Color online). Phenomenon of avoided crossing expected from
  a lattice calculation of the static $NN$ energy as a function of the
  distance, $E_{NN}(r) = 2 M_N + V_{NN} (r)$ in the Hilbert space
  spanned by NN and multi-pion states, ${\cal H} = {\cal H}_{NN} \oplus
  {\cal H}_{NN\pi} \oplus {\cal H}_{NN 2\pi} \oplus \dots $. Anytime a
  crossing with a $n \pi $ threshold occurs, $V_{NN} (r_n) = n m_\pi$,
  the system chooses the minimum energy state with the pions at
  rest. The potential $V_{NN} (r)$ with the repulsive core corresponds
  to the envelop of all the different branches.}
\label{fig:ENNstatic}
\end{figure}

On general grounds one should expect pion production when
$V_{NN}(r_\pi) \ge m_\pi$ in which case the potential should develop
an imaginary part as the system becomes unstable against the decay $NN
\to NN\pi$ and pions will eventually be radiated.  This feature is
precluded in the quenched approximation where particle creation is
suppressed. Alternatively one could, in addition to just $NN$ states,
implement $NN\pi$ configurations (for instance taking $J_\pi (x) \sim
\bar q(x) \vec \tau \gamma_5 q(x) $ located at $x=0$), still within
the quenched approximation, in which case a coupled channel
Hamiltonian spanning trial Hilbert space ${\cal H}= {\cal H}_{NN}
\oplus {\cal H}_{NN\pi}$ and incorporating, besides $NN \to NN $ and
$NN\pi \to NN\pi $ diagonal elements, the $NN \to NN\pi$
transition. Schematically, one has 
\begin{eqnarray}
{\cal H} (r) - 2M_N = \left( \begin{matrix}  V_{NN,NN}(r) & V_{NN,NN \pi}(r) 
 \\ V_{NN,NN \pi}(r)  &  m_\pi + V_{NN \pi,NN \pi}(r)  \end{matrix} \right) \, . 
\end{eqnarray}
The eigenvalues of this Hamiltonian may be denoted by $E_- (r) < E_+
(r) $ and clearly one has the variational relation $E_- (r) \le
V_{NN,NN}(r)+ 2M_N$. Under these circumstances, an avoided crossing pattern,
familiar from molecular physics in the Born-Oppenheimer
approximation~\cite{landau1965quantum} will occur as a function of the
separation distance $r$, choosing the lower energy $NN\pi$ branch
below the pion-production distances. The situation is sketched in
Fig.~\ref{fig:ENNstatic} for the case of avoided crossings with
several pions assuming a small channel mixing and vanishing $V_{NN ,
  n\pi ; NN , n\pi} (r)$ for simplicity.

Of course, this argument can be generalized when further inelastic
channels, such as $N\Delta$ or $\Delta \Delta$ are open. From a
variational point of view, we always have
\begin{eqnarray}
{\rm min} E_n (r) \le V_{NN,NN}(r) + 2 M_N \, , 
\end{eqnarray}
so that we can regard the repulsive core found in QCD lattice
calculations~\cite{Aoki:2011ep,Aoki:2013tba} as an upper bound of the
true static energy within the restricted two nucleon Hilbert space,
${\cal H}_{NN}$, and {\it not} as a genuine feature of the NN
interaction. This observation is one of our main motivations for the
present paper. 

\subsection{High energy NN analysis}

While the discussion of the core shape and details had some impact in
nuclear matter and the nuclear equation of state one can also
undertake a similar analysis more directly within a free nucleons
scattering context. Definitely, a hard or soft repulsive core located
at a short distance such as $a_c=0.5$fm can only be resolved and
isolated from the rest of the potential when the relative de Broglie
reduced wavelength becomes of much smaller size, i.e. $1 /p_{\rm CM}
\gg a_c$. This means going up to LAB energies $T_{\rm LAB} = 2 p_{\rm
  CM}^2 /M_N \gg 340 {\rm MeV}$, above pion production threshold and
well beyond the traditional domain of NN potentials used in nuclear
physics for nuclear structure and nuclear reactions. Fortunately, there
exist by now abundant data of pp and np scattering permitting a model
independent partial wave analysis going up to $T_{\rm LAB}^{\rm max}=3
{\rm GeV}$~\cite{LechanoineLeLuc:1993be}, i.e. $p_{\rm CM}^{\rm
  max}=1.2 {\rm GeV}$ allowing a direct reconstruction of scattering
amplitudes~\cite{Ball:1998an,Bystricky:1998rh,Ball:1998jj} from
complete sets of experiments. In general, phase shifts can only be
considered observables when a complete set of measurements
(differential cross sections, polarization asymmetries, etc.) for a
fixed energy has been measured.  In the late 70's and until today
extensive fits have been undertaken in the region below the onset of
diffractive scattering. The NN PWA at energies well above the pion
production threshold has a also long history and a good example of
subsequent upgrades is represented by the series of works conducted by
Arndt and
collaborators~\cite{Arndt:1982ep,Arndt:1986jb,Arndt:1992kz,Arndt:1997if,Arndt:2000xc,Arndt:2007qn}
(see also the GWU database~\cite{SAID}). The most recent GWU fit
\cite{Workman:2016ysf} is based on a
parameterization~\cite{Arndt:1986jb} with a total number of 147
parameters and fitting up to a maximum of all $J=7$ partial wave
amplitudes (phases and inelasticities), up to 3 GeV deals with a large
body of $25362$-pp data (with $\chi^2= 48780.934$) and $13033$-np data
(with $\chi^2= 26261.000$), which is sufficient for our considerations
here.

\subsection{Short distance correlations}

In many respects the short distance aspects of the NN interaction are
relevant for nuclear physics at intermediate energies. The typical
example is provided by short distance correlations, where the
traditionally accepted repulsive core should become more
visible. Experimentally, such effects might be ``seen'' in two nucleon
knock out experiments (e,e',NN) and they would be responsible for the
high-momentum pair distribution~\cite{Shneor:2007tu}. Large scale
calculations generate such distributions using a real NN potential to
solve the nuclear many body
problem~\cite{Feldmeier:2011qy,Vanhalst:2012ur,Alvioli:2013qyz,Wiringa:2013ala}. Of
course, at relative CM momenta where $p= \sqrt{M \Delta} \sim 2k_F
\sim 600 {\rm MeV}$ and $\Delta=M_\Delta-M_N= 0.297 {\rm GeV}$ the
$\Delta-N$-splitting, the inelasticity becomes large and the
interaction in free space cannot be fixed by a real and energy
independent NN potential. This corresponds to back-to-back collisions
of particles on the surface of the Fermi sphere and provides a natural
cut-off for these momentum distributions stemming from real NN
potentials.  In a recent study~\cite{RuizSimo:2016vsh} this issue has
been analyzed and it has been found that for CM momenta below $p
\lesssim \sqrt{M\Delta}$ ($T_{\rm LAB} \lesssim 2 \Delta$) there is no
need for a nuclear repulsive core. In fact, the main contribution
stems from the mid-range attractive part. In scattering experiments
particles are on-shell and that means that energy and momentum are
related. In finite nuclei, where particles are off-shell, we can
witness short-distance properties, i.e., high momentum states but
keeping their energy inside the nucleus small.  Therefore, that means
that an energy independent NN interaction might be fixed directly from
the high-momentum distribution rather than from NN scattering. In any
case, the traditional evidence of the repulsive core based on short
distance correlations should be revised when inelasticities are taken
into account.

\subsection{Relativity and inelasticity}

In NN scattering at the high energies under consideration, $T_{\rm
  LAB} \lesssim 3 {\rm GeV}$ we have three essential features: Spin
degrees of freedom, opening of inelastic channels and relativity.  At
these energies and since $\sqrt{s} \sim 2 M_N + n m_\pi$ up to a
maximum 8 pions (among other things) can be produced.  The full
multichannel calculation directly embodying $NN \to NN + n \pi$
transitions is prohibitive and has never been carried out. In reality
much less pions are produced in average $\bar n \sim 2$ since the
largest contributions to the inelastic cross section stem from
resonance production and decay, say $NN \to N \Delta \to NN \pi$ or
$NN \to 2 \Delta \to N N 2 \pi$, etc. triggered by peripheral pion
exchange.

The general field theoretical approach would require a coupled channel
Bethe-Salpeter equation, where the kernel would ultimately be
determined phenomenologically from the NN scattering data (see
e.g. \cite{Eyser:2003zw}). Under these circumstances the effort of
solving the Bethe-Salpeter equation may be sidestepped by a much
simpler procedure, namely the invariant mass
framework~\cite{Allen:2000xy} which corresponds to solve
\begin{eqnarray}
{\cal M}^2 \Psi  \equiv 4 ( -\nabla^2 + M_N^2 ) \Psi + M_N
V \, \Psi = s \, \Psi \, , 
\end{eqnarray} 
with $s$ the standard Mandelstam variable and to identify the
relativistic and non-relativistic CM momenta $p_{\rm CM}^2 =
s/4-M_N^2$ yielding an equivalent Schr\"odinger equation with a
potential $V$.  We will incorporate inelastic absorption via an
optical (complex) potential
\begin{eqnarray}
V(r,s)= {\rm Re} V(r,s)+
i {\rm Im} V(r,s) \, , 
\end{eqnarray}
by appealing to the standard Feshbach
justification~\cite{Feshbach:1958nx,Feshbach:1962ut} of separating the
Hilbert space into elastic and inelastic sectors corresponding to the
$P$ and $Q$ orthogonal projectors. Field theoretical approaches
assuming the conjectured double spectral representation of the
Mandelstam type provide a link to this optical potential as well as
its analytic properties in the
$s$-variable~\cite{cornwall1962mandelstam,omnes1965optical}. Optical
potential approaches have already been proposed in the past in this
energy range and an early implementation of the optical potential in
NN scattering within the partial wave expansion was carried out in
Ref.~\cite{Ueda:1973zz}.  A more microscopic description involves
explicitly $N \Delta$ inelastic channels (see e.g.
Ref.~\cite{Eyser:2003zw} and references therein).  A relativistic
complex multirank separable potential of the neutron-proton system was
proposed in Ref.~\cite{Bondarenko:2008mm,Bondarenko:2011za}. The
approach we use here furnishing both relativity and inelasticity has
already been exploited in the much higher energy range covering from
ISR up to LHC, $\sqrt{s} = 25-7000 {\rm
  GeV}$~\cite{Arriola:2016bxa,RuizArriola:2016ihz}.

In this paper we will approach the problem by suitably adapting the
coarse graining idea presented in a series of papers to this new
inelastic
situation~\cite{Entem:2007jg,NavarroPerez:2011fm,NavarroPerez:2012qr,Perez:2013cza,NavarroPerez:2012qf,Perez:2013jpa}. As
already mentioned, at the partial waves level most of the evidence
about the core comes from the S-waves, so we will restrict our study
to this simple case.

\section{Short wavelength fluctuations}
\label{sec:fluctuations}

\subsection{Inverse scattering}

In order to motivate our subsequent discussion, let us approach the
problem from an inverse scattering point of view (for a review see
e.g.  \cite{chadaninverse}). Such inverse scattering methods have also
been extended within NN scattering in the inelastic
regime~\cite{Leeb:1985zz,Khokhlov:2004mf}, and to some extent
represent a model independent determination of the underlying
interaction. By all means these approaches require a complete
description of the phase shift from threshold up to infinity, i.e.  $0
< p_{\rm CM} < \infty$. In practice, any truncation at high energies,
say $p_{\rm CM}=\Lambda$, generates a short wavelength ambiguity
$\Delta r = \hbar /\Lambda$ which we can regard as a short distance
fluctuation, since finer resolutions will effectively become
physically irrelevant.  This should not be a problem for a real
potential where only elastic scattering may take place, and since
Levinson's theorem guarantees the phase-shift to go to zero at high
energies. For a complex potential where inelastic channels are open
the situation may be quite different in practice, since, firstly, we
do not have a Levinson's theorem and, secondly, the optical potential
may not even go to zero at high energies~\footnote{For instance, in
  Ref.~\cite{Eyser:2003zw} the modeling nucleon-nucleon scattering
  above 1-GeV has been addressed signalling a gradual failure of the
  traditional one boson exchange (OBE) picture which needs rescaling
  $1/s$ of the OBE potential.}.

Inverse scattering methods also allow a geometric glimpse into the NN
inelastic hole at several distances and
energies~\cite{Geramb:1998ps,Funk:2001ph} when a repulsive core is
assumed for lab energies below 3 GeV, although it has also been
recognized that this solution is not unique~\cite{Khokhlov:2004mf}.
In the case of Ref.~\cite{Knyr:2006ka}, where relativistic optical
potentials on the basis of the Moscow potential and lower phase shifts
for nucleon scattering at laboratory energies up to 3 GeV were
considered it was found that there was no core representation of the
inverted interaction for energies above 1GeV.  Actually, in
Ref.~\cite{Khokhlov:2005vp} one can identify the oscillations of the
resulting local potential. These short wavelength
fluctuations/oscillations are inherent to the maximum energy or CM
momentum $\Lambda$ being fixed for the phase shift.  The local
projection method based on close ideas also produces very similar
oscillations~\cite{Wendt:2012fs}.

\subsection{Coarse graining}
\label{sec:coarse}

Under these circumstances we will invoke from a Wilsonian point of
view the coarse graining of the interaction down to the shortest
resolution scale in the problem. This is based on the reasonable
expectation that wavelength fluctuations shorter than the smallest de
Broglie wavelength, which determine the maximal resolution given by
$\Delta r \sim 1/p$, are unobservable. For a potential $V(r)$ with a
typical range $r_c$ this simply means taking a grid of points which is
chosen for convenience to be equidistant, $r_n = n \Delta r $, and
using the potential at the grid points $V(r_n)=V_n $ as the fitting
parameters themselves. On the other hand, the long range part of the
potential will be taken to be given by One Pion Exchange (OPE) above
$r_c=3 {\rm fm}$~\footnote{The choice of this boundary is not
  arbitrary; it has been motivated from comprehensive PWA at energies
  about pion production threshold (see below).}, thus we will have
\begin{eqnarray}
V(r) = V_{\rm Short} (r) \theta (r_c-r) + V_{\rm OPE} (r) \theta (r-r_c) \, , 
\end{eqnarray}
where in the $^1S_0$ pn channel 
\begin{eqnarray}
V_{\rm OPE}^{^1S_0 } (r) = -3 f^2 \frac{e^{-m_\pi r}}{r} \, , 
\end{eqnarray}
Of course, there are many possible ways to coarse grain the inner
component of the interaction $V_{\rm Short}(r)$ and a particularly
simple one has been to take a sum of delta-shells as initially
proposed by Avil\'es in 1973~\cite{Aviles:1973ee} and reanalyzed in
Ref.~\cite{Entem:2007jg} (see
\cite{NavarroPerez:2011fm,NavarroPerez:2012qr,Perez:2013cza} for
pedagogical reviews) and pursued in latter studies
\cite{NavarroPerez:2012qf} (a comprehensive mathematical analysis can
be consulted in Ref.~\cite{albeverio2013spherical}) yielding the most
accurate determination of $f^2=0.0763(1)$~\cite{Perez:2016aol} when
charged pions and many other effects are considered. Here, we will not
attempt this very high accuracy and take the more conventional value
$f^2=0.075$ and $m_\pi = 140 {\rm MeV}$ for simplicity. In this paper,
we will take also piecewise square-well potential for definiteness
and, because it looks more intuitive, will use it for the main
presentation. Of course, none of our main results depends on the
particular regularization and we will provide also results for the
delta-shell coarse graining. Our notation will be as follows:
\begin{eqnarray}
V_{\rm GR}^I (r)= 
V_{\rm SW} (r) & \equiv& \sum_{i=1}^N  V_i^{\rm SW} \theta(r_{i-1} < r \le r_i) \, , \label{eq:sw-pot} \\
V_{\rm GR}^{II} (r)= 
V_{\rm DS} (r) & \equiv & \sum_{i=1}^N \Delta r \, V_i^{\rm DS} \delta (r-r_i) \, , \label{eq:ds-pot} \\
& r_n = n \Delta r \, . 
\end{eqnarray}
Note that roughly we expect $V_i^{\rm SW} \approx V_i^{\rm DS} $. The
corresponding S-wave Schr\"odinger equation has to be solved with the
boundary conditions
\begin{eqnarray}
u(0)=0 \, , \qquad u(r) \to \sin \left( p r + \delta (p) \right) \, . 
\end{eqnarray}
From the continuity of the function and the (dis)continuity of the
derivative it is then straightforward to obtain a recurrence relation
whence the total accumulated phase-shift may be obtained. The
approximation involved in Eq.~(\ref{eq:sw-pot}) and
Eq.~(\ref{eq:ds-pot}) can be regarded as simple integration methods
for the Schr\"odinger equation, i.e.  given a potential $V(r)$ we can
take grid points $r_n= n \Delta r$ and $V_n \equiv V(r_n)$. We
obviously expect the {\it fine} graining limit, i.e. $\Delta r \to
0$ and $N \to \infty $ with $r_c= N \Delta r$ fixed we get an
arbitrary good solution to the wave function. We give below the
corresponding discretized formulas for the two cases.

\subsubsection{Square well}

For a sequence of square well potentials $U_1 , \dots U_N $ with $U_i
\equiv 2\mu V(r_i) $, the solution can be written piecewise
\begin{eqnarray}
u_0 (r) &=& \sin( K_1 r + \delta_0) \, , \qquad 0 < r < r_1 \, , \\
u_1 (r) &=& \sin( K_2 r + \delta_1) \, ,  \qquad r_1 < r < r_2 \, , \\
\dots &&  \\
u_N (r) &=& \sin( K_{N+1} r + \delta_N) \, ,  \qquad r_N < r \, , 
\end{eqnarray}
where $K_n = \sqrt{-U_n + k^2}$.  We take $\delta_0=0$ and $\delta_n$
is the accumulated phase-shift due the sequence of square wells, so
that $\delta(k)= \delta_N(k)$. Matching the log-derivative at every 
$r_i$ we get the recursion relation
\begin{eqnarray}
\cot \delta_{n+1} = \frac{A_n+ B_n \cot \delta_n }{C_n + D_n \cot \delta_n} \, , 
\label{eq:rec-sw}
\end{eqnarray}
where 
\begin{eqnarray}
A_n &=& K_{n+1} \cot \left(K_n r_n\right)-K_n \cot \left(K_{n+1} r_n\right) \, , \\ 
B_n &=& K_n \cot \left(K_n r_n\right) \cot \left(K_{n+1} r_n\right)+K_{n+1} \, , \\
C_n &=& K_{n+1} \cot \left(K_n r_n\right) \cot \left(K_{n+1} r_n\right)+K_n \, , \\
D_n &=& K_{n+1} \cot \left(K_{n+1} r_n\right)-K_n \cot \left(K_n r_n\right) \, . 
\end{eqnarray}

\subsubsection{Delta-shells}

For the sequence of delta-shells potentials with coefficients $ \Delta r
U_1 , \dots \Delta r U_N $ the solution can similarly be written
piecewise
\begin{eqnarray}
u_0 (r) &=& \sin( k r + \delta_0) \, , \qquad 0 < r < r_1 \, , \\
u_1 (r) &=& \sin( k r + \delta_1) \, , \qquad r_1 < r < r_2 \, , \\
\dots &&  \\
u_N (r) &=& \sin( k r + \delta_N) \, ,  \qquad r_N < r \, . 
\end{eqnarray}
As before, we take $\delta_0=0$ and $\delta_n$ is the accumulated
phase-shift due the sequence of square wells.  Matching the
log-derivative at any $r_n$ we get the recursion relation
\begin{eqnarray}
k \cot \left( k r_n + \delta_{n+1} \right)- k \cot \left( k r_n +
\delta_{n} \right) = U_n \Delta r
\end{eqnarray}

\subsection{Counting parameters for the optical potential}

In contrast to a conventional integration method, where the
integration step $\Delta r $ is an auxiliary parameter which should be
removed $\Delta r \to 0$ based on the accuracy of the wave function,
in the coarse graining approach $\Delta r \sim 1/p_{\rm CM}$ becomes a
physical parameter which only goes to zero when the scattering energy
goes to infinity and the precision is dictated by the experimentally
measurable phase shifts.  One of the advantages of this point of view
is that we avoid using specific functional forms which possibly
correlate different points in the potential and hence introduce a bias
in the analysis. A crucial question is to know how many {\it
  independent} fitting parameters $V(r_n)$ are needed to produce a
good fit to the scattering data going to a maximum CM momentum p. The
simple answer for the S-wave is just $N \sim r_c /\Delta r = p r_c $.
(for higher partial waves and coupled channels see
e.g. \cite{Perez:2013cza}).~\footnote{If one disregards the OPE tail
  above $r_c = 3 {\rm fm}$, the delta-shell method provides in
  momentum space~\cite{Entem:2007jg} a multirank separable potential
  very much in the spirit of the proposal in
  Ref.~\cite{Bondarenko:2008mm,Bondarenko:2011za}. Our coarse grain
  argument does in fact foresee {\it a priori} the rank of the
  interaction; it coincides with the number of grid points $N$. This
  result holds also when the inelasticity is taken into account (see
  below).}

This very simple idea underlies comprehensive bench-marking NN studies
about pion production threshold undertaking a complete
PWA~\cite{Entem:2007jg,NavarroPerez:2011fm,NavarroPerez:2012qr,Perez:2013cza,NavarroPerez:2012qf}.
In this work we will extend the idea to the case of interest of a
complex optical potentials where the imaginary component of the
potential takes care of the absorption. Definitely, for this analysis
to be competitive the number of parameters must be controllable {\it a
  priori}. The coarse graining involves also the inelasticity hole,
which we assume to be of the order of the traditional nuclear core, as
will be verified below. Therefore, in our analysis in S-waves and up
to the highest energies we consider the inelasticity as point-like
since $\Delta r \sim a_c$, an assumption which will be justified
below.

Resolving the core structure in more detail requires going to higher
energies, where there is no PWA so that the separate contribution of
the S-wave could be analyzed. Therefore, either more complete data
will be needed or other methods can be used. For instance, above 3
GeV Regge behavior sets in and we refer to recent works undertaking
such an analysis~\cite{Sibirtsev:2009cz,Ford:2012dg}. At much higher
energies such as those measured at ISR ($\sqrt{s} \sim 25-50 \, {\rm
  GeV}$) and LHC ($\sqrt{s} \sim 2000-14000 \, {\rm GeV}$) and
sufficiently large momentum transfers one has $\Delta r \ll a_c $ and
the structure of the inelasticity hole can be pinned down more
accurately assuming spin independent
interactions~\cite{Arriola:2016bxa,RuizArriola:2016ihz}.

\begin{figure*}[t]
\begin{center}
\includegraphics[scale =1.2]{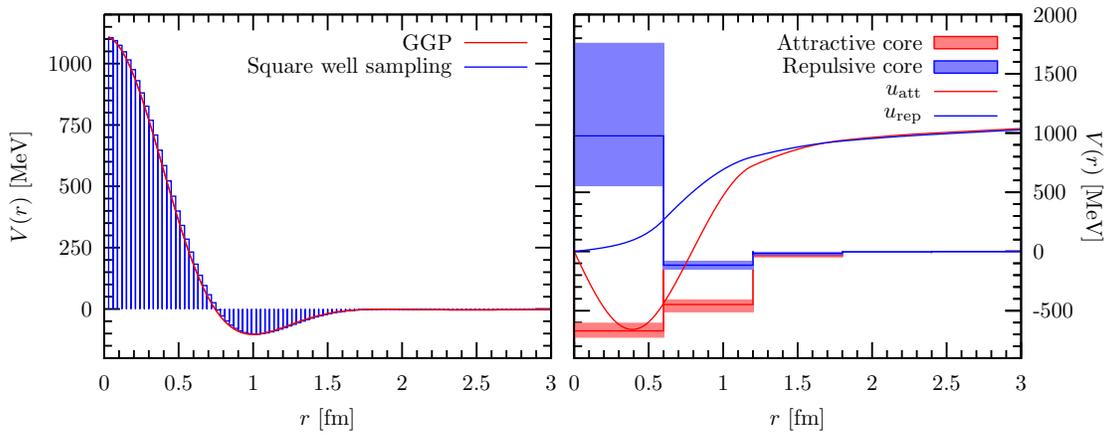}
\end{center}
\caption{(Color online). (Left panel) Granada smooth Gaussian
  potential in the $^1S_0$ channel~\cite{Perez:2014yla} and the
  sequence of square wells with heights $V_n$ sampling the values of
  the original potential $V_n \equiv V(r_n)$. The total number of
  points is $N=40$. (Right panel): (Short distance) Attractive and
  repulsive coarse grained sequence of square wells fitted to the
  Granada phase-shifts up to 300 MeV LAB energy. Uncertainties stem
  from the phase-shifts. The total number of points is $N=4$. We also
  plot the corresponding zero energy wave function  in arbitrary units.}
\label{Fig:pots300}
\end{figure*}

\begin{figure*}[t]
\begin{center}
\includegraphics[scale =1.2]{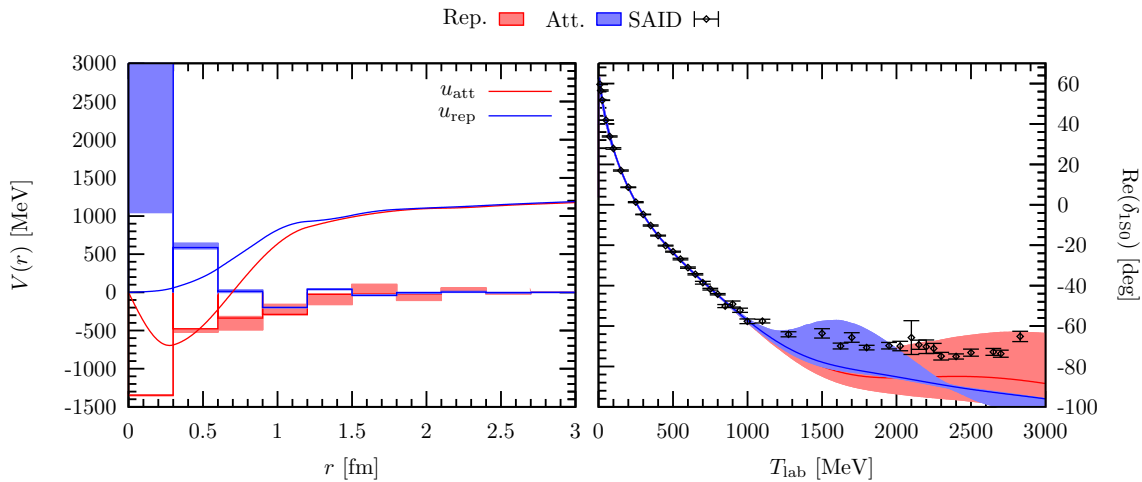}
\end{center}
	\caption{(Left panel): (Short distance) Attractive and
          repulsive coarse grained sequence of square wells fitted to
          the real part of the SAID $^1S_0 $ np phase-shifts up to 1 GeV LAB
          energy. Uncertainties stem from the phase-shifts. The total
          number of points is $N=10$. We also plot the corresponding
          zero energy wave function in arbitrary units. (Right panel)
          Comparison of the fits with the SAID database up to 3 GeV.}
\label{pot-pozos-1000}
\end{figure*}

\section{Elastic coarse graining revisited}
\label{sec:elastic}

In previous works by the Granada group~\cite{Perez:2013jpa}
extensive use of delta-shell based coarse graining potentials has been
made in order to carry out the most comprehensive NN fit to date up to
350 MeV.  This was done for a total of $N_{\rm Dat} =6713$ np + pp
scattering observables slightly above the pion production threshold
with a very high statistical quality, $\chi^2/\nu=1.04$ and total
number of $N_{\rm Par}=46$ independent parameters providing some
confidence on the method. In this section we revisit the fit for the
$^1S_0$ channel by using instead a piecewise square-well potential. As
already said, and in harmony with previous findings we will assume OPE
interaction above $r_c= 3{\rm fm}$.

\subsection{Fit to 300 MeV}

Let us consider for definiteness the Gaussian-OPE potential obtained
in Ref.~\cite{Perez:2014yla} in the $^1S_0$ channel, which as we see
from Fig.~\ref{Fig:pots300} presents a repulsive core starting at 0.6
fm. We may integrate the differential equation by using any standard
procedure, such as Numerov's method (see e.g. Ref.~\cite{Oset:1985tb}
and references therein) where the precision is on the determination of
the wave function as a primary quantity. In order to reproduce the
phases with the smallest uncertainty quoted in
Ref.~\cite{Perez:2014yla} and taking a maximum integration distance of
$R=5$fm we need $M=2^{14}=16384$ integration points. For our purposes
this high pointwise accuracy in the wave function is not strictly
necessary, as the wave function is not an observable, and we are
merely interested in determining the physically measurable
phase-shift.

Therefore, we may regard the piecewise solutions described in
Section~\ref{sec:coarse} as integration methods and study the accuracy
with respect to the phase-shift and {\it not} with respect to the wave
function. For instance, if we use a piecewise square well potential as
an integration method using Eq.~(\ref{eq:rec-sw}) we get that with
about 40 wells which values are given by the potential $ U_i= U(r_i)$
prove sufficient, a result shown in Fig.~\ref{Fig:pots300}. This
corresponds to the {\it fine graining} point of view for a {\it
  prescribed} potential.

However, the potential was obtained from a fit to data, and in this
case to the phase shift. The question now is on how many {\it fitting}
wells, $U_1^{\rm Fit}, \dots U_N^{\rm Fit}$, separated by $\Delta r$
are needed to reproduce the same phase shift to a certain accuracy up
to a maximal energy value which we take here to be 300 MeV. As already
anticipated, the answer is $N=5$, a much smaller value than with the
fine graining case. Our results are again shown in
Fig.~\ref{Fig:pots300} and as we see there are two possible solutions,
corresponding to an {\it attractive} and {\it repulsive} potential. We
will refer to them as A and R respectively and the numerical values
can be looked up in table \ref{Parametros-1S0-300}. For completeness
the delta-shell fitting values are also presented in
Table~\ref{Parametros-1S0-deltas-300}. The existence of two solutions
in itself is not surprising as it reflects and illustrates in the
coarse grained framework the well-known ambiguities of the inverse
scattering problem~\cite{Leeb:1985zz}. The physical reason is that the
corresponding wavelength does not sample the short distance region
with sufficient resolution. As a matter of fact we will see that these
two solutions depart from each other at energies higher than those
used in the fitting procedure.

\begin{table}[H]
\begin{center}
\begin{tabular}{|c|c|c|}
\hline
$r_i$(fm) &  $V_i^{\rm SW}$(MeV) - A &  $V_i^{\rm SW}$(MeV) - R  \\
\hline
$0.6$ & $-661(48)$ & $ 1175(536)$ \\
$1.2$ & $-459(43) $ & $ -119(30)$ \\
$1.8$ & $ -35(8) $ & $  -16(9)$ \\
$2.4$ & $  -5(1) $ & $   -3(3)$ \\
$3.0$ & $ -0.1(0.4)$ & $   -0.4(0.8)$ \\
\hline
\end{tabular}
\caption{Values of the squared well potentials at the given sampling
  points obtained fitting up to $T_{\rm lab}=300$ MeV for both the
  attractive (A) and repulsive (R) solutions.  They are obtained from
  a fit to a number of $N_d=10$ data with a corresponding
  $\chi^2=2.9$. The central values and uncertainties of the parameters
  were obtained as expected values and standard deviations from set of
  fits to synthetic data, generated from the experimental
  uncertainties.  }
\label{Parametros-1S0-300}
\end{center}
\end{table}

    \begin{table}[H]
    \begin{center}
    \begin{tabular}{|c|c|c|}
    \hline
$r_i$(fm) &  $V_i^{\rm DS}$(MeV) - A &  $V_i^{\rm DS}$(MeV) - R  \\
    \hline
    $0.6$ & $-265(4) $ & $ 127(20)$ \\
    $1.2$ & $-293(6) $ & $ -66(5)$ \\
    $1.8$ & $ -64(7) $ & $   -3(3)$ \\
    $2.4$ & $  -2(2) $ & $   -3(2)$ \\
    $3.0$ & $  -2(4) $ & $    1(3)$ \\
    \hline
    \end{tabular}
    \caption{Same as Table~\ref{Parametros-1S0-300} but for delta-shells.}
    \label{Parametros-1S0-deltas-300}
    \end{center}
    \end{table}

\subsection{Fit to 1 GeV}

It is natural to think that by increasing the energy we will be able
to resolve more accurately the short distance region.  More
specifically, we might be able to pin down the nature of the
nuclear core as well as better discriminating between both solutions A
and R found before and eventually ruling out one of both solutions. As
said, above pion production threshold the potential must reflect the
inelasticity, but we also expect this effect to be located at short
distances as suggested by the small inelastic cross
section. Therefore, in a first attempt we will take $T_{\rm LAB}<
1$GeV and fit just the real part of the phase shift with a real and
energy independent potential.

In our analysis at higher energies we will profit from the PWA carried
out by several groups in the past and will use the GWU
database~\cite{SAID} for definiteness. Moreover, we will restrict
ourselves to the simplest case of the most important S-waves since our
main purpose is to merely show that the coarse graining works at much
higher energies also when inelasticities are included. Moreover,
S-waves have the smallest possible impact parameter $b=1/(2p)$ sensing
the core region.  In a further publication we will extend the analysis
to higher partial waves.

Let us briefly review the basic idea and count the number of necessary
parameters. The maximum resolution corresponds to the shortest de
Broglie wavelength, which is $\lambda=0.3 {\rm fm}$. On the other
hand, we assume as it was done in the low energy analysis that from
$r_c \ge 3 {\rm fm}$ the only contribution is due to One Pion
Exchange.  Thus, we have to sample $r_c/\Delta r= 10$ points.

The result of the fit is now presented in Fig.~\ref{pot-pozos-1000}.
We get $\chi^2/\nu = 1.6$ in both cases, which shows that inelasticity
must be taken into account even if only the real part of the phase
shift is fitted (see below). By all means when we are dealing with a
real potential where the inelasticity is small but non-zero, we may
wonder what is the uncertainty in the potential associated to
this. Numerical values with their uncertainties for the repulsive and
attractive core in the SW and DS can be seen in Tables~\ref{Parametros-1S0-1000} and \ref{Parametros-1S0-deltas-1000} respectively.

\begin{table}[H]
\begin{center}
\begin{tabular}{|c|c|c|}
\hline
$r_i$(fm) &  $V_i^{\rm SW}$(MeV) - A &  $V_i^{\rm SW}$(MeV) - R  \\
\hline
$0.3$  & $-1352(34)$ & $ 6851(1553)$ \\
$0.6$  & $ -494(28)$ & $  360(285)$ \\
$0.9$  & $ -395(84)$ & $   66(64)$ \\
$1.2$  & $ -234(72)$ & $ -195(6)$ \\
$1.5$  & $  -81(70)$ & $   17(29)$ \\
$1.8$  & $   32(66)$ & $  -11(31)$ \\
$2.1$  & $  -56(42)$ & $  -26(23)$ \\
$2.4$  & $   32(25)$ & $   17(14)$ \\
$2.7$  & $  -13(10)$ & $   -8(7)$ \\
$3.0$ & $    2(3)$ & $    0(2)$ \\
\hline
\end{tabular}
\caption{Same as Table \ref{Parametros-1S0-300} 
but fitting up to $T_{\rm lab}=1000$ MeV. In this case  $\chi^2=33.9$ and $N_d=24$.}
\label{Parametros-1S0-1000}
\end{center}
\end{table}

    \begin{table}[H]
    \begin{center}
    \begin{tabular}{|c|c|c|}
    \hline
$r_i$(fm) &  $V_i^{\rm DS}$(MeV) - A &  $V_i^{\rm DS}$(MeV) - R  \\
    \hline
    $0.3$  & $-741(20)$   & $ 4643(2000)$ \\
    $0.6$  & $ -430(111)$ & $   533(97) $ \\
    $0.9$  & $ -377(21)$   & $  -233(20) $ \\
    $1.2$  & $ -143(4)$     & $      -4(10)$ \\
    $1.5$  & $  -25(3)$     & $    -34(5)   $ \\
    $1.8$  & $  -13(6)$     & $     -6(5)  $ \\
    $2.1$  & $  -13(4)$     & $     -8(4)   $ \\
    $2.4$  & $   11(3)$     & $      6(4)   $ \\
    $2.7$  & $   -7(2)$     & $     -5(3)  $ \\
    $3.0$ & $   3(3)$     & $       1(3)  $ \\
    \hline
    \end{tabular}
    \caption{Same as Table~\ref{Parametros-1S0-1000} but considering
      delta-shells. The chi-square obtained in this case is
      $\chi^2=33.0$.}
    \label{Parametros-1S0-deltas-1000}
    \end{center}
    \end{table}

\begin{figure*}[ht]
\begin{center}
\includegraphics[scale =1.1]{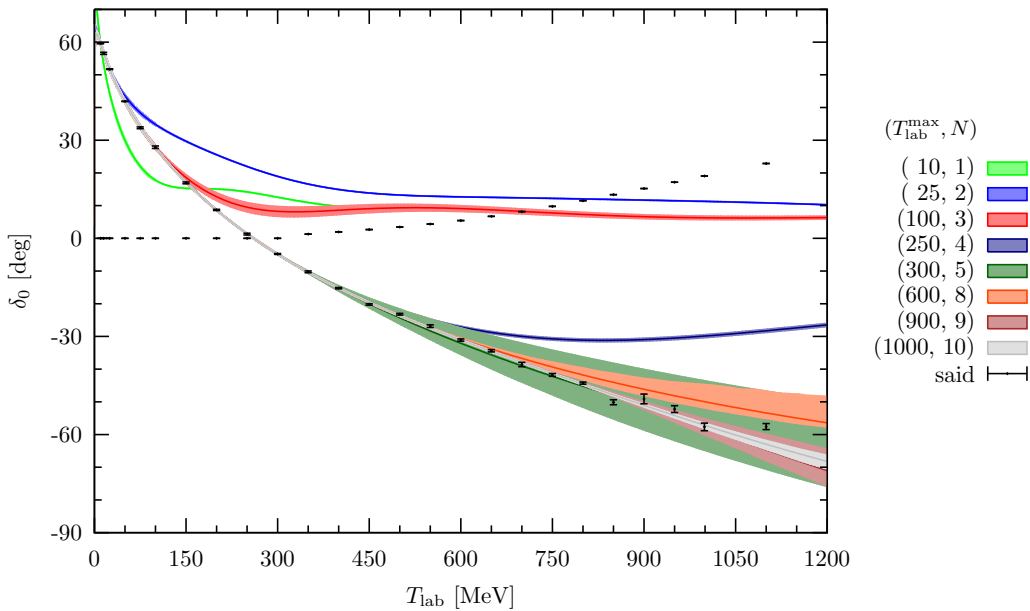}
\end{center}
	\caption{np phase shifts (in degrees) in the $^1S_0$ channel
          as a function of the LAB energy (in MeV). We show gradual
          fits in the repulsive core case to increasingly high energies to
          the real part of the phase-shift with an increasing number
          of square wells with the corresponding extrapolated
          errorbands. For illustration we also show the imaginary part
          of the phase shift (non-fitted).}
\label{pot-evolution}
\end{figure*}

\subsection{Quality of fits}

The quality of fits can be tested in a number of ways, including the
$\chi^2$-test (see e.g. \cite{Perez:2015pea}). The verification of
these tests ensures uncertainty propagation.  For completeness, we
also show in Table~\ref{Normality-of-the-residuals} the moments test
for the lowest moments, and as we see the test validates error
propagation within the expected fluctuations due to a finite number of
fitting data. In all quoted results the central values and
uncertainties of the parameters were obtained as expectation values and
standard deviations from set of fits to synthetic data, generated from
the experimental uncertainties using the bootstrap
method~\cite{Perez:2014jsa}.

    \begin{table*}
    \begin{center}
    \begin{tabular}{|c|cccc|cccc|c|}
    \hline
    $ $ &  \multicolumn{4}{c}{$T^{\rm max}_{\rm lab}=0.3$ GeV} & \multicolumn{4}{c}{$T^{\rm max}_{\rm lab}=1$ GeV} & $ $ \\[0.3 mm]
    $ $ & SW - A & SW - R &  DS - A & DS - R & SW - A & SW - R &  DS  - A&DS  - R & $\mathcal{N}(0,1)$\\
    \hline
    $\mu_3$  & $-0.37 $  & $-0.18$ & $-0.08$  & $-0.07$ & $-0.11$   &  $-0.11$    &    $-0.17$   &   $-0.15 $&   $0\pm 1.223 $\\
    $\mu_4$  & $ 2.65$  & $ 2.63$ & $ 2.47$  & $ 2.53$  & $ 2.89$   &  $ 2.89$    &    $ 2.87$   &    $2.89  $&   $3\pm3.093 $\\
    $\mu_5$  & $-2.89$  & $-1.64$ & $-0.87$  & $-0.79$  & $-1.24$   & $-1.24$    &    $-1.61$   &    $-1.50 $&   $0\pm 9.681$\\
    $\mu_6$  & $ 9.83$  & $ 9.23$ & $ 8.12$  & $ 8.49$  & $ 11.5$   & $ 11.5$     &    $ 11.4$   &    $11.5   $&   $15\pm 31.827 $\\
    $\mu_7$  & $-15.0$  & $-8.53$ & $-4.46$  & $-4.11$  & $-8.80$   & $-8.80$     &    $-10.7$   &    $-10.2 $&   $0\pm 115.842 $\\
    $\mu_8$  & $ 39.6$  & $ 34.1$ & $ 28.0$  & $ 29.6$  & $ 50.6$   & $ 50.6$     &    $50.4 $   &    $ 50.8 $&   $105\pm 434.016 $\\
    \hline
    \end{tabular}
    \caption{Test of the normality of the residuals relative to the fits involving deltas-shells and squared wells. Note that the different moments $\mu_r$ must be compared with the standard gaussian distribution $\mathcal{N}(0,1)$ moments in the last column.}
    \label{Normality-of-the-residuals}
    \end{center}
    \end{table*}

\begin{figure*}
\centering
\epsfig{figure=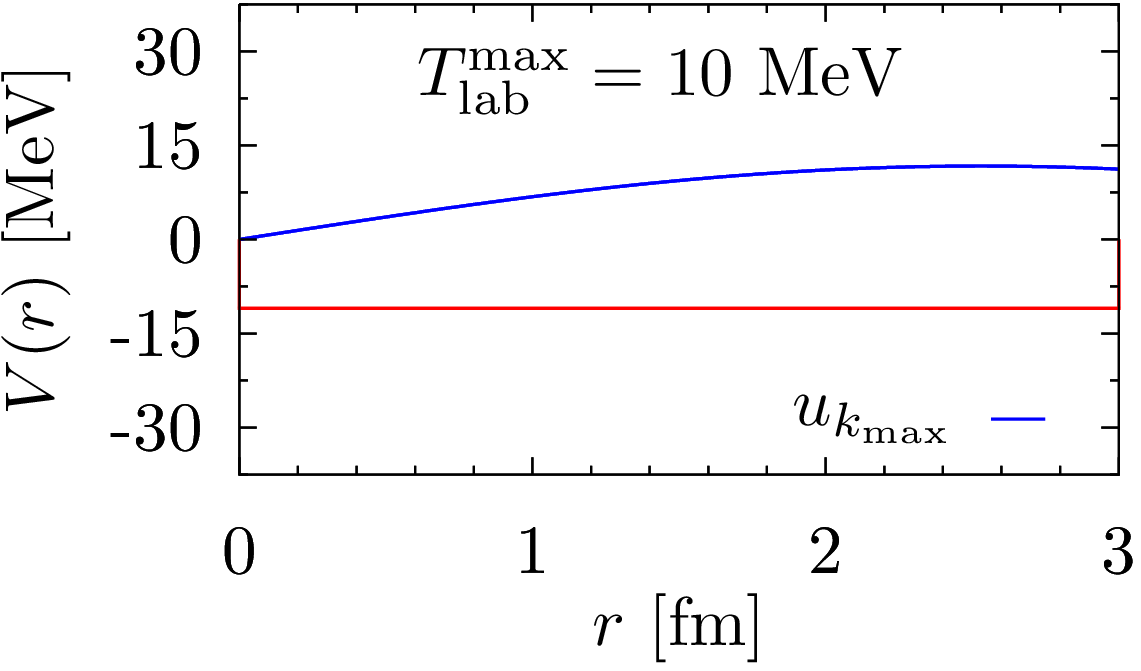,width=4cm,height=4cm}
\epsfig{figure=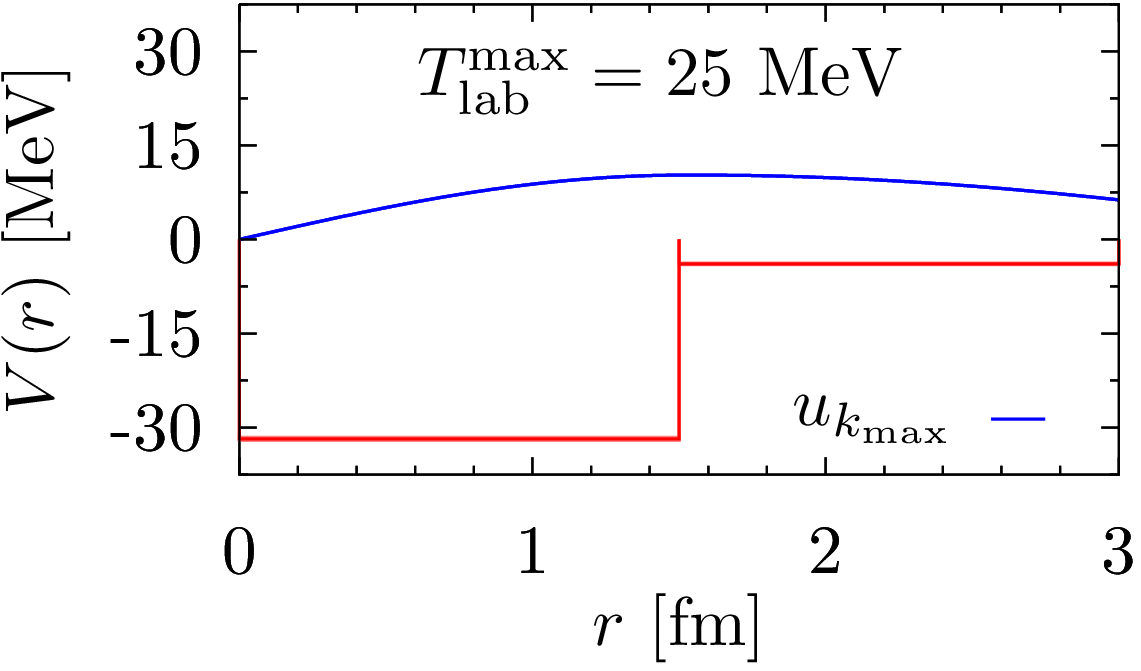,width=4cm,height=4cm}
\epsfig{figure=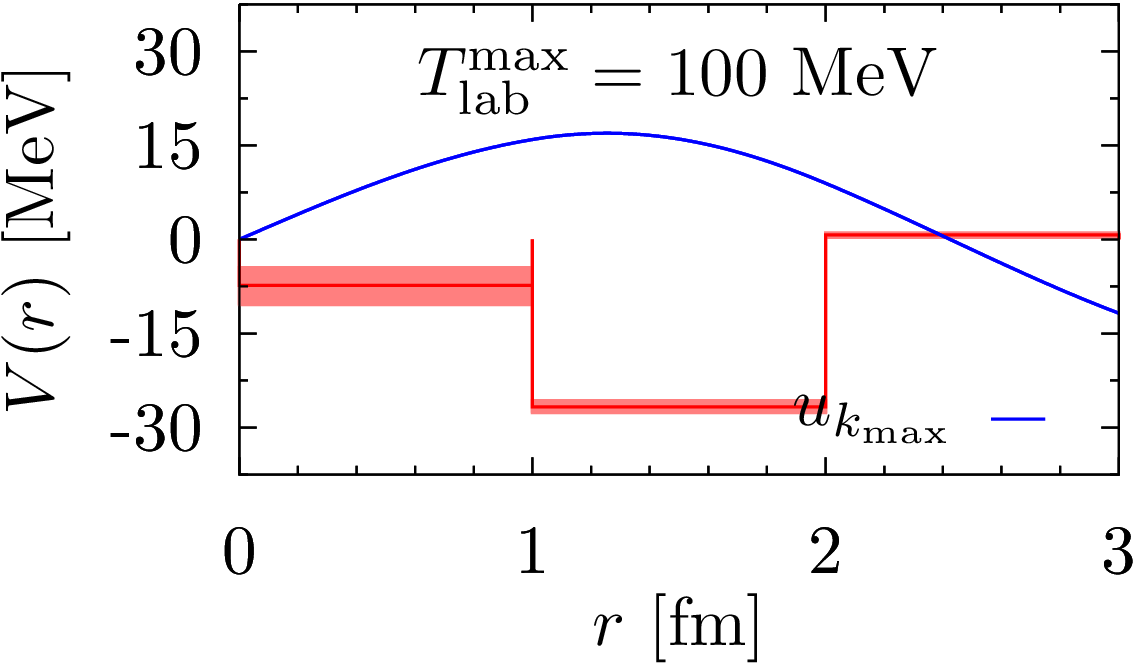,width=4cm,height=4cm}
\epsfig{figure=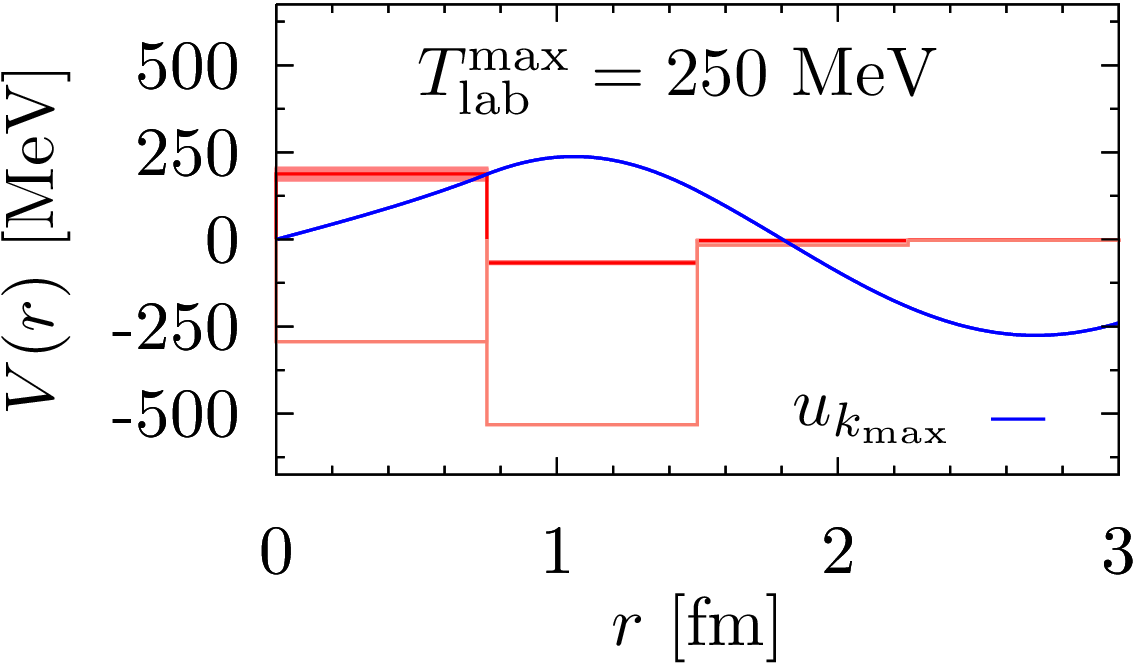,width=4cm,height=4cm}\\
\epsfig{figure=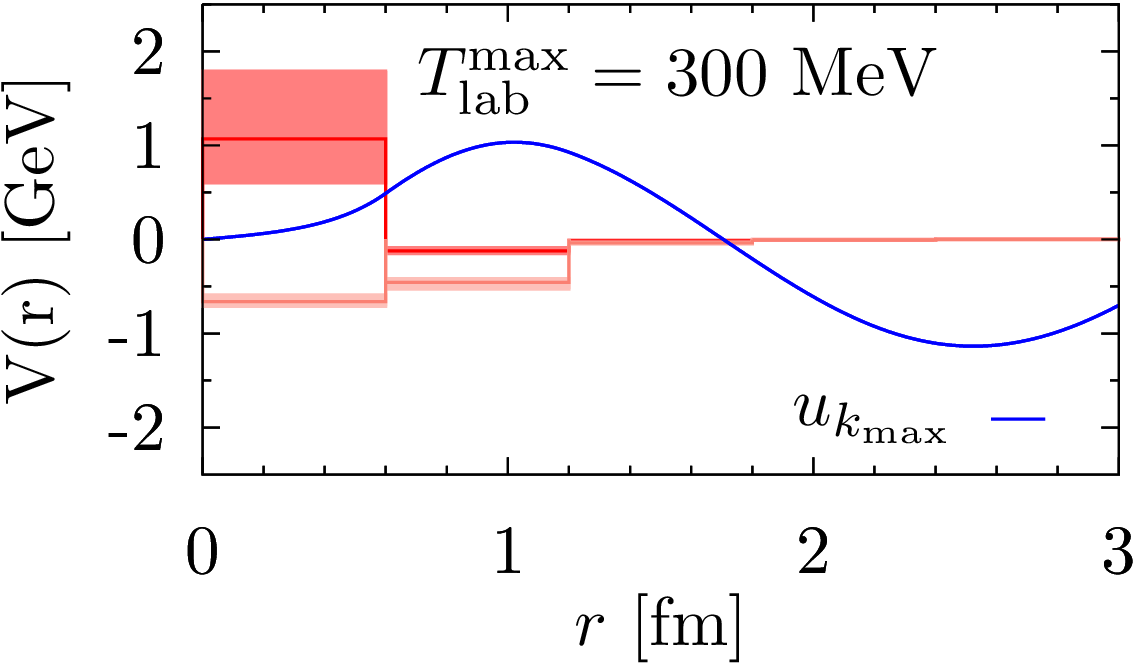,width=4cm,height=4cm}
\epsfig{figure=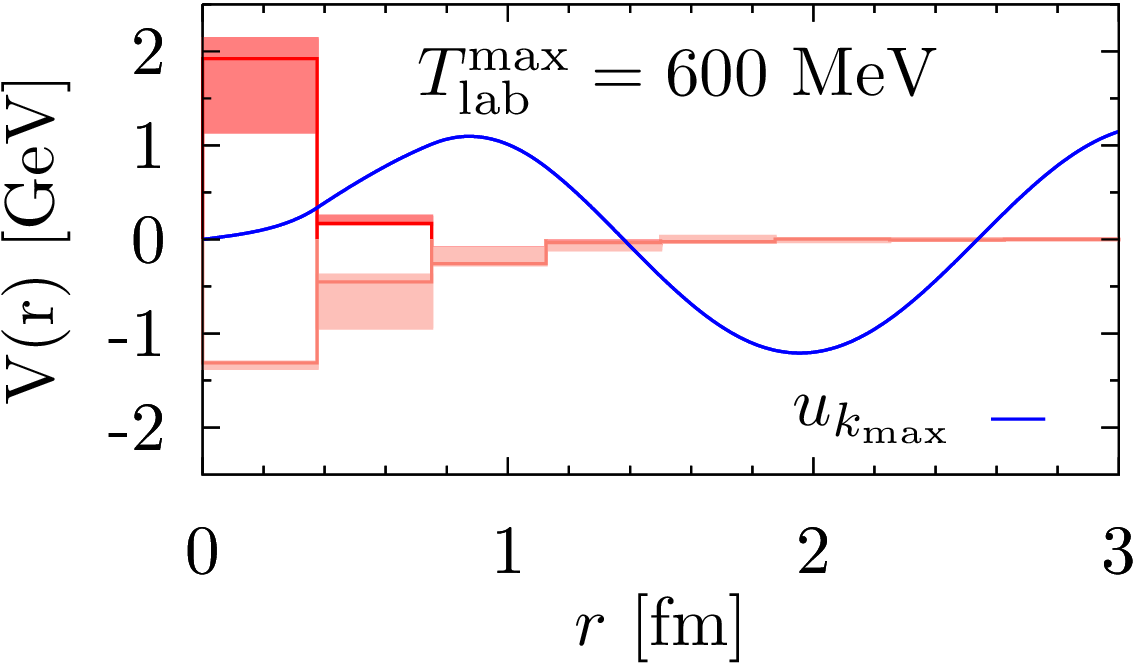,width=4cm,height=4cm}
\epsfig{figure=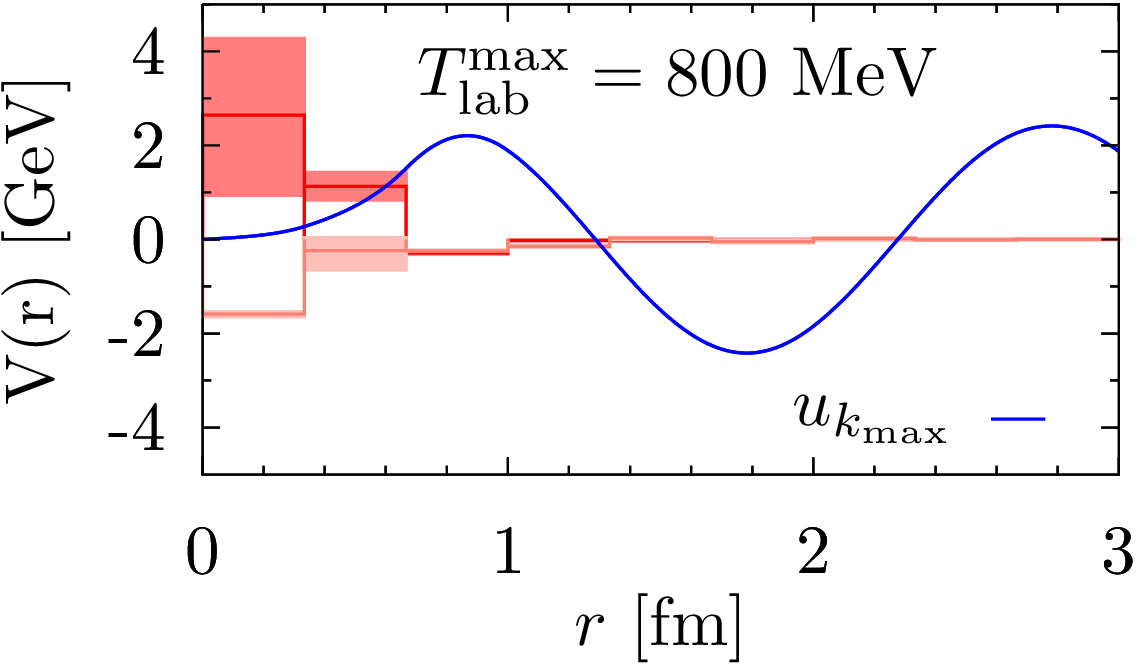,width=4cm,height=4cm}
\epsfig{figure=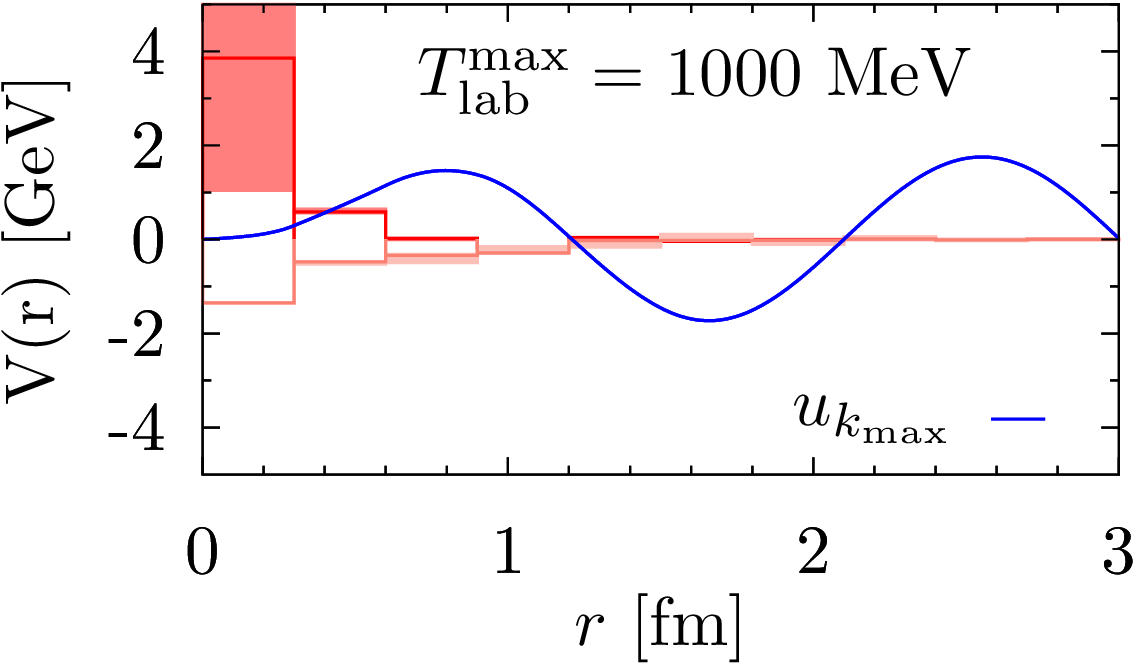,width=4cm,height=4cm}
\caption{The evolution of the different square well potentials with
  the maximal fitting LAB energy. For illustration we also plot the
  maximum energy wave function in arbitrary units. The bifurcation
  between the repulsive and attractive core occurs at about $250$MeV.
} 
\label{fig:coarsed-wells}
\end{figure*}

\subsection{Evolution of potentials as a function of the fitting energy}

In order to illustrate the coarse graining idea, we show in
Fig.~\ref{pot-evolution} the evolution of the phases as the maximal
fitting energy is gradually increased in the repulsive core
case. Essentially we just take as a guide that $\Delta r \sim 1/p_{\rm
  CM, max}$ and the number of square wells is approximately given by
$N \sim r_c/\Delta r$. The errors are propagated from the fitting
results beyond the fitting range. As we see, already with $N=5$ the
prediction band contains the SAID data for which we plot both the real
and the imaginary part of the phase-shift.  A close study on the the
evolution of the phases was carried out in a previous
work~\cite{RuizSimo:2016vsh} taking the AV18
potential~\cite{Wiringa:1994wb} as a reference interaction which
reproduces {\it a fortiori} the scattering data. There the coarse
graining of the interaction with delta-shells was studied with similar
conclusions.  Here, besides using the square wells we also analyze the
impact of phase-shift uncertainties stemming from the fits to the
experimental data in the analysis. The evolution of the square well
potentials as a function of the maximal fitting energy together with
the corresponding wave function with the shortest wavelength are
depicted in Fig.~\ref{fig:coarsed-wells} illustrating the meaning of
coarse graining in this particular case.

\section{Coarse grained Optical Potentials}
\label{sec:optical}

We come now to the fit up to 3 GeV which, as mentioned above, is the
largest LAB energy where a complete PWA has been undertaken and hence
the particularly interesting $^1S_0$-wave has been extracted.  Of
course, at these high energies one may produce up to 7 pions among
other particles, a circumstance that is beyond any comprehensive
theoretical analysis at present due to the large number of coupled
channels and multiparticles states. Therefore, and in line with
previous developments and the discussion above we will use an optical
potential. At the energies where the inelastic cross section is
sizeable, relativistic corrections start playing a role. As mentioned
above we will implement relativity invoking the invariant mass
framework~\cite{Allen:2000xy} and already exploited in recent high
energy studies~\cite{Arriola:2016bxa}.

\begin{figure}
\begin{center}
\epsfig{figure=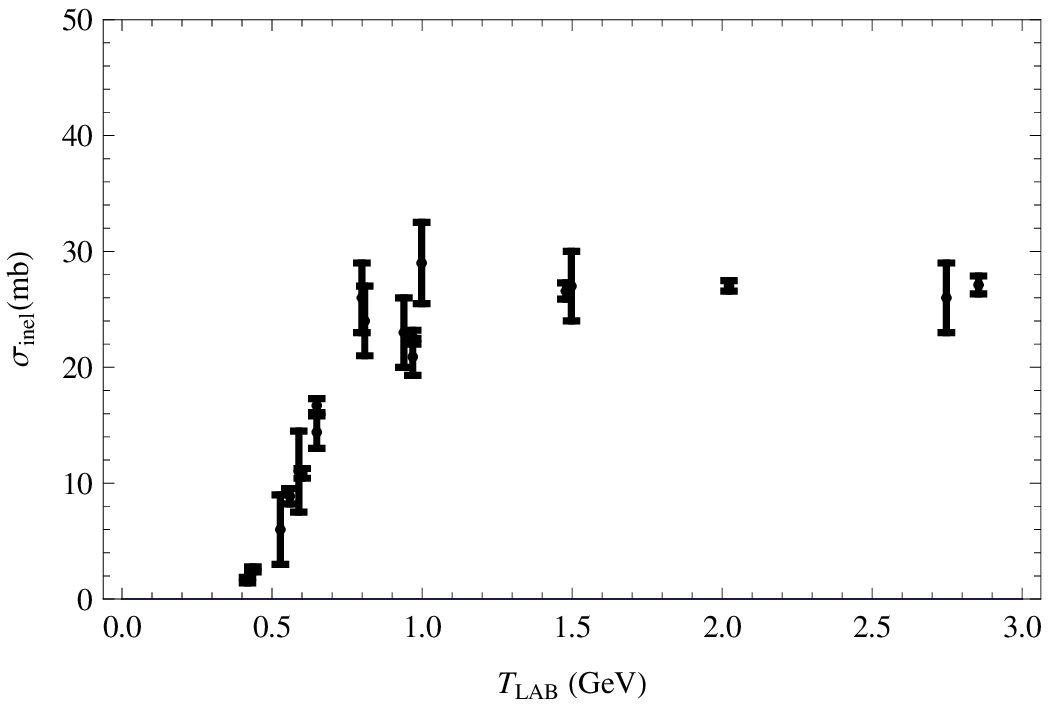,width=8cm,height=6cm} 
\end{center}
\caption{Neutron-Proton Inelastic cross section as a function of the
  LAB energy\cite{LechanoineLeLuc:1993be}. The $\Delta$ and $2 \Delta$
  resonance regions happen at $T_{\rm LAB}= 0.64 {\rm GeV}$ and
  $T_{\rm LAB}= 1.36 {\rm GeV}$ respectively.}
\label{sigma-inel} 
\end{figure}

\begin{figure}
\begin{center}
\epsfig{figure=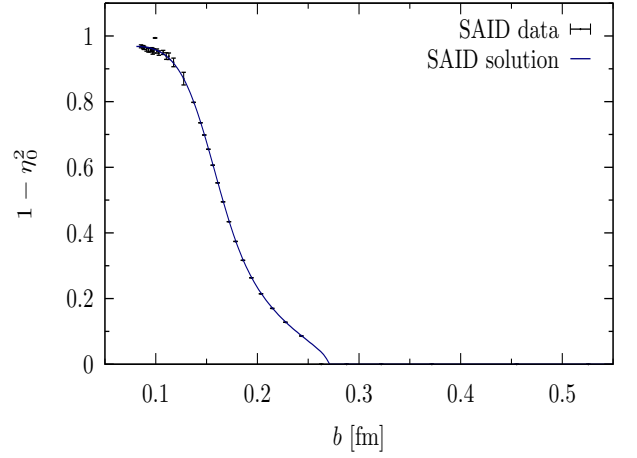,width=8cm,height=6cm} 
\end{center}
\caption{Neutron-Proton inelasticity profile for the $^1S_0$ channel
  $1-\eta_0(b)^2$ as a function of the impact parameter $ b =
  (l+1/2)/p$ for $l=0$ from the SAID database.}
\label{blackdisk} 
\end{figure}

\begin{figure}[ht]
\begin{center}
\includegraphics[scale =2]{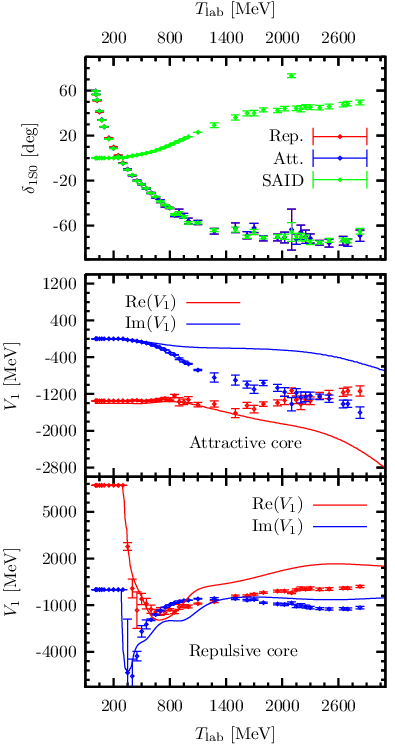}
\end{center}
\caption{ Fit to the real and imaginary part of the $^1S_0 $ np SAID
  phase-shift as a function of the LAB energy up to 3 GeV for both the
  attractive and repulsive core cases (Top panel).  Energy dependence
  of the ${\rm Re} V_1$ and ${\rm Im} V_1$ values keeping the
  remaining parameters to the energy independent values for the
  attractive core (middle panel) and the repulsive core (bottom
  panel). Continuous lines represent using the SAID solution and
  discretized lines represent direct determination from scattering
  data.}
\label{inelasticity}
\end{figure}

In general the coarse grained optical potential will be complex, and
a pertinent question is what is the range of the
inelasticity. We address this issue by noting that the inelastic cross
section jumps from $\sigma_{\rm in} = 1 {\rm mb}$ to $\sigma_{\rm in}
= 28-32 {\rm mb}$ in the CM range between
$\sqrt{s}=2.3-2.7$GeV~\cite{Ryan:1971rv} (for a review see
e.g. \cite{LechanoineLeLuc:1993be} and references therein), which
corresponds to nucleon resonance excitations $NN \to N \Delta$ or $NN
\to \Delta \Delta$ for $ M_N + M_\Delta \lesssim \sqrt{s} \lesssim 2
M_\Delta$, i.e. $\sqrt{M \Delta} \lesssim p \lesssim \sqrt{2 M
  \Delta}$ with $\Delta = M_\Delta-M_N$, and remains almost constant
$\pm 1 {\rm mb}$ up to 1000 GeV. We show as an illustration
$\sigma_{\rm in}$ in Fig.~\ref{sigma-inel} up to $3 {\rm GeV}$ LAB
energy. In the energy range we are concerned with $\sigma_{\rm in}
\sim \sigma_{\rm el}$ so that the absorbing black disk picture
holds. That means that the size of the hole is about 1fm.  There are
at least two ways to represent the absorptive S matrix, either as a
complex phase-shift or as a real phase-shift and an inelasticity.  The
relation between both is
\begin{eqnarray}
S_l(p) = \eta_l (p) e^{ 2 i \delta_l (p)}=e^{-2\rho_l (p)}e^{2i\delta_l (p)}
\end{eqnarray}
On the other hand, the inelastic cross section can be written as
a sum over partial waves, 
\begin{eqnarray}
\sigma_{\rm in} = \frac{\pi}{p^2}\sum_{l=0}^\infty (2 l+1) \left[ 1-\eta_{l}(p)^2 \right]   
\end{eqnarray}
In the impact parameter representation where we can set $(l+1/2) =
bp$, the S-wave involves the shortest impact parameter $b_{\rm
  min}=1/(2p)=1/\sqrt{s-4 M^2}$ and at this level we can see from
Fig.~\ref{blackdisk} that, $1-\eta_0 (p)^2$ vanishes for $b$ larger
than the coarse graining scale $ \Delta r = 0.3 {\rm fm}$.  Thus, we
will assume {\it just one} complex and energy dependent fitting
parameter for the innermost potential well, i.e.  at $r=r_1 \equiv
\Delta r$,
\begin{eqnarray}
V(r_1,p) &=& {\rm Re} V(r_1,p) + i  {\rm Im} V(r_1,p) \equiv {\rm Re} V_1 + i  {\rm Im} V_1 
\end{eqnarray}
and keep the remaining $N-1$ points fixed and energy independent to
the values of the previous fit at $T_{\rm LAB}=1 {\rm GeV}$.  The
previous recursion relations, Eq.~(\ref{eq:rec-sw}), hold also here by just
replacing $\delta_l(p) \to \delta_l(p) + i \rho_l(p) $.  The
inelasticity and energy dependence at $r=r_1$ already improves the
previous fit from $\chi^2/\nu=1.6$ to $\chi^2/\nu=0.7$. 

\begin{figure}[ht]
\begin{center}
\includegraphics[scale =0.7]{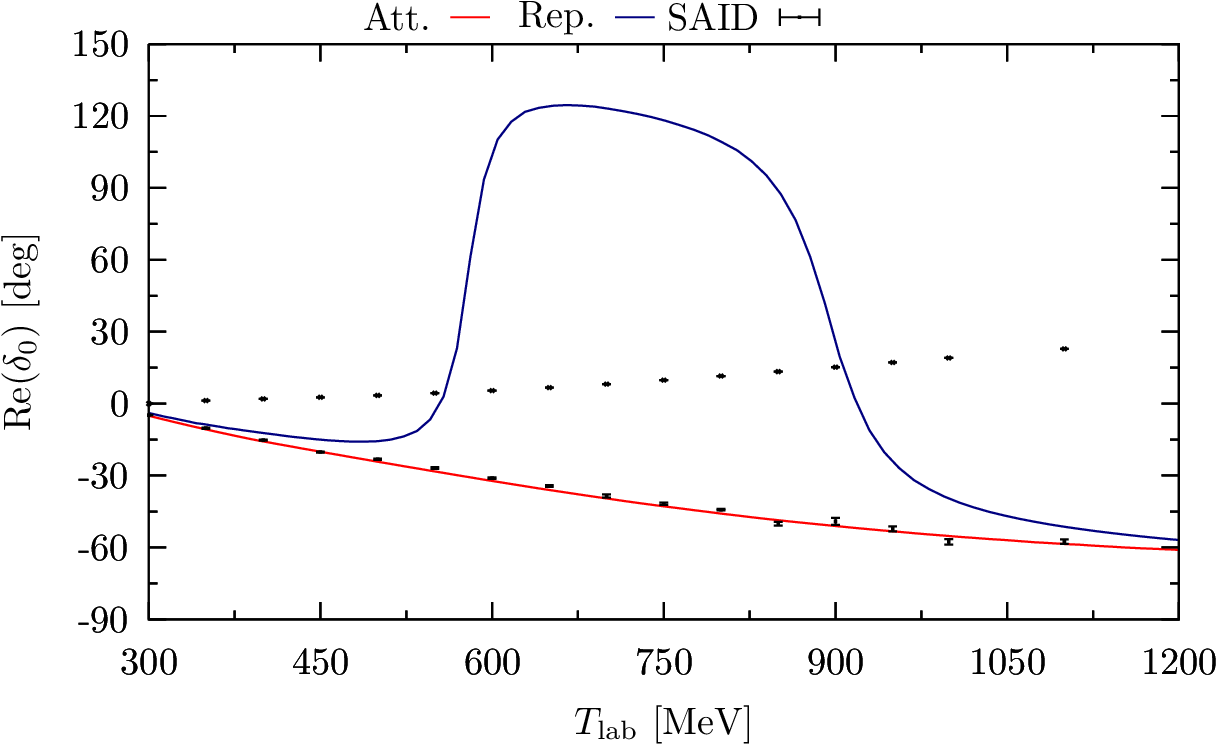}
\end{center}
\caption{Effect of switching off the imaginary part of the potential
  ${\rm Im} V_1=0$ keeping the remaining parameters in the $^1S_0$
  np-phase-shift in the attractive and repulsive core scenarios.}
\label{switch-off}
\end{figure}

Our procedure is as follows: for any energy value we fit the real and
imaginary part of the phase shift following the strategy of taking
just one complex and energy dependent parameter at the innermost
sampling point. This will generate a discretized energy dependence of
${\rm Re} V(r_1,p_\alpha)$ and ${\rm Im} V(r_1,p_\alpha)$, with
$T_{\rm LAB, \alpha}= 2 p_{\rm CM,\alpha}^2/M_N $ and $\alpha=1, \dots
, N_{\rm Dat}$. In order to produce a smooth continuous curve we have
also used the SAID solution which we find that it produces worse fits
than the set of discretized energies.

The dependence of the innermost real and imaginary parts is depicted
in Fig.~\ref{inelasticity} while the remaining components are kept
energy independent from the previous low energy fit. As we can see the
behavior is quite different. While in the attractive core case both
${\rm Re} V_1$ and ${\rm Im} V_1$ exhibit a rather smooth energy
dependence, the repulsive core scenario presents a rapid and sudden
change already in the region where the inelasticity is small. It is
interesting to see what is the effect of switching off the inelastic
contribution {\it without} refitting parameters. As we see in
Fig.~\ref{switch-off}, the effect is dramatic in the standard
repulsive core scenario and very mild in the attractive core case.
Therefore, we can conclude that the inelastic contribution behaves
truly as a perturbative effect in the structural core scenario.

\section{Discussion and outlook}
\label{sec:sc}

\subsection{The structural core and cluster models}

Our finding for the repulsive branch echoes the result of G.E. Brown,
60 years ago~\cite{brown1958proton} namely a rapid and non-adiabatic
transition from a repulsive core to an extremely absorbing disk.  The
underlying and microscopic reason why this sudden change might happen
has never been clarified to our knowledge.

\begin{figure*}[ht]
\begin{center}
\includegraphics[scale =1.3]{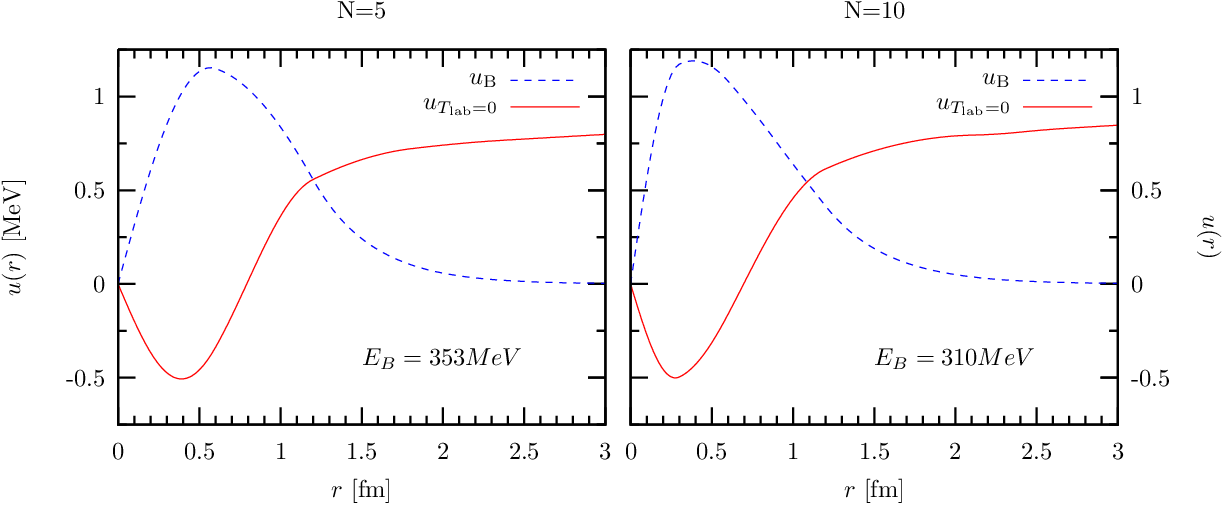}
\end{center}
\caption{Zero energy wave function and spurious Pauli-forbidden bound state 
when the fit is done with $n=5$ (left) and $n=10$ (right) square wells 
in the $^1S_0$
  np-phase-shift in the attractive core scenario.}
\label{fig:spurious}
\end{figure*}

Unlike the repulsive branch, the attractive branch presents a deeply
bound state at $E=-350 {\rm MeV}$ already at the lowest LAB energy
fits considered. This implies in particular that the zero energy wave
function must have a node due to the oscillation theorem. In
Fig.~\ref{Fig:pots300} (right) we show the zeroth energy wave function
for both the attractive and repulsive branches in the 350 LAB energy
fit case. As we see, they both vanish at short distances and feature
the difference between the repulsive core, where the wave function
vanishes below the core radius, and the structural core, where the
wave function simply oscillates below the node. This pattern was
already encountered in the OBE analysis and the main difference was
taking an unnaturally large SU(3) violating $g_{\omega NN} \sim 20 $
coupling constant or a natural one $g_{\omega NN} = 3 g_{\rho NN} \sim
10 $~\cite{Cordon:2009pj}. Here we confirm the trend when short
distances are really probed by reproducing high energy scattering.
The zero energy wave function pattern does not change much when the
fitting energy is raised to $1 {\rm GeV}$, as can be seen in
Fig.~\ref{pot-pozos-1000}. In Fig.~\ref{fig:spurious} we show the
shape of the deeply bound state for two maximal fitting energies
corresponding to fit $n=5$ and $n=10$ wells.

Of course, this deeply bound state, while not influencing strongly the
scattering, it would provide a very negative contribution to the
binding energy in finite nuclei by just placing pairs of protons and
neutrons in $^1S_0$ channel with the relative spurious bound state
wave function. While at first sight this may seem an unsurmountable
problem for the structural core scenario, in what follows we argue why
this spurious deeply bound state could and should indeed be removed
from the potential having no direct impact on the nuclear structure
calculation.

The argument is based on old considerations on the structural
core~\cite{Otsuki:1965yk,TAMAGAKI:1967zz,Otsuki:1969qu} and they are
most easily and beautifully exemplified by the discussion on $^8$Be
made of two $\alpha$-clusters. Within a quark cluster model scheme for
the nucleon, these nodes in the zero energy wave function are expected
as a consequence of the Pauli principle of the constituent quarks,
where not all quarks belonging to different nucleons are
simultaneously exchanged, and hence is a more general requirement than
the Pauli principle for nucleons.

One should note that the exact location of the node is searched within
the standard cluster Gaussian wave function approach because for these
states the CM can be easily extracted. A nice interpretation of the
phenomenon can be given within the Saito orthogonality condition model
which he analyzed thoroughly for two $\alpha$
clusters~\cite{saito1969interaction,saito1977chapter} (for
comprehensive reviews see e.g.
\cite{tang1978resonating,Friedrich:1981ad}) and we remind here briefly
adapted to the nucleon-nucleon situation. Due to the total
antisymmetry of the wave function, the interaction term between
clusters contains a direct (Hartree) and an exchange (Fock) term,
where the quarks inside one nucleon are exchanged with the quarks in
another nucleon. For a local quark-quark interaction the direct
nucleon-nucleon term is also local whereas the exchange term is
nonlocal, and this is a short distance contribution which becomes
negligible when the nucleons are at distances larger than their
size. Thus, at large non-overlapping distances the direct and local
term survives and one can work within a Hartree approximation.  Within
a Hartree-Fock framework Saito found that the Fock term had the effect
of generating a node in the wave function when working in the Hartree
approximation, i.e., when just the direct term is kept and the
exchange term is ignored. Thus, there is no direct node-less zero
energy wave function, since it is Pauli forbidden, and thus the
corresponding deeply bound state which one finds in the Hartree
approximation is spurious !. The Fock term generates the extra node of
the wave function and prevents the nodeless wave function.

Our coarse grained potential corresponds to just using the direct
term. Therefore, the node we find in the zero energy wave function is
just a manifestation of the composite character of the nucleons when
they are handled as if they were elementary. These views have been
advocated in several papers~\cite{Neudatchin:1975zz,Neudachin:1977vt}
motivating the series of Moscow potentials~\cite{Kukulin:1992vd}.

\subsection{Extension to higher partial waves}

The fact that coarse graining is working at these high energies in the
case under study here with just S-wave is encouraging. It would thus
be quite interesting to extend the analysis to higher partial waves,
and actually it should be possible to make a complete coarse grained
PWA using the delta-shells regularization extending previous work at
about pion production threshold. Following the argument of a previous
work~\cite{Perez:2013cza} we may count the total number of parameters
by setting the maximal angular momentum $l_{\rm max} = p_{\rm max}
r_c$ , and using $p_{\rm max} \Delta r \sim 1 $.  On the other hand,
for any $l \le l_{\rm max}$ we take $ l(l+1)/r_{\rm min} = p_{\rm
  max}^2$. Thus we get using the $2 \times 2$ spin-isospin degeneracy
factor corresponding to neutron and proton states with spins up and
down
\begin{eqnarray}
N_{\rm par} = 4 \sum_{l=0}^{l_{\rm max}} \sum_{n} \theta \left( p^2
- \frac{l(l+1)}{r_n^2} \right) \sim 4 \frac12 (p_{\rm max} r_c)^2
\end{eqnarray}
For the studied case here where $r_c=3 {\rm fm}$ and $T_{\rm LAB} = 3
{\rm GeV}$ , i.e. $p_{\rm CM}= 1.26 {\rm GeV}$ one gets $N_{\rm par}
\sim 700 $, or equivalently $N_{\rm par}^{pp} = N_{\rm par}^{nn} \sim
175 $ and $N_{\rm par}^{pn} \sim 350 $. These are large numbers of
parameters, but not very different from the ones needed in the
comprehensive most recent SAID np fit~\cite{Workman:2016ysf} based on
a parameterization~\cite{Arndt:1986jb} with a total number of 147
parameters. It remains to be seen if these extra parameters might
perhaps improve the quality of these fits since they describe
$25362$-pp data with $\chi^2= 48780.934$ and $13033$-np data with
$\chi^2= 26261.000$.

\subsection{Implications for nuclear physics}

Fitting NN scattering data is not only a possible way to represent the
data, but also the fitted potential is meant to be used in a nuclear
structure calculation. While this seems most obvious, the well known
complexities of the nuclear many-body problem require making a choice
on the characteristics features of the potential itself. For instance,
the mere concept of a repulsive core requires considering a local
potential. This particular form is best suited for MonteCarlo type
calculations and underlies the Argonne potentials
saga~\cite{Lagaris:1981mm,Wiringa:1984tg,Wiringa:1994wb}.  Nonlocal or
velocity dependent potentials do not exhibit the repulsive core so
explicitly~\footnote{It should be reminded that one can always perform
  a unitary phase-equivalent transformation from one form to another~
  (see e.g. \cite{Neff:2015xda} and references therein).}  and are
usually handled by other techniques, such as no-core shell model.

On the other hand, if we want to access to the short distance region
in finite nuclei by say, knock out processes (e,e',NN), we must fix
the NN interaction going to high enough energies where both
inelasticities and relativity ought to be important ingredients of the
calculation.  The resulting interaction, such as the one determined in
the present paper, will thus become complex and energy dependent, and
it is unclear at present how to use such an interaction in
conventional Nuclear Structure calculations from an {\it ab initio}
point of view where the main assumption is to take real and energy
independent potentials. We hope to address these issues in the future.

\subsection{Implications for Hadronic Physics}

The fact that the coarse graining approach works for NN scattering in
a regime where relativistic and inelastic effects become important
suggests extending the method to other hadronic reactions under
similar operating conditions such as $\pi\pi$ scattering up to
$\sqrt{s}=1.4 {\rm GeV}$~\cite{jacobo2017}. Within such a context the
methods based in analyticity, dispersion relations and crossing are
currently considered to be, besides QCD, the most rigorous
framework~(See e.g. \cite{Ananthanarayan:2000ht, GarciaMartin:2011cn}
and references therein). We stress that such an approach is based on
the validity of the double spectral representation of the four-point
function conjectured by Mandelstam. It is noteworthy that under this
same assumption the optical potential of the form used in the present
paper was derived many years
ago~\cite{cornwall1962mandelstam,omnes1965optical}; analytic
properties for $V(r,s)$ in the $s$-variable in the form of a
dispersion relation at {\it fixed} relative distance $r$ were deduced.
Due to their linear character dispersion relations would be preserved
under the coarse graining operation at the grid points for $V(r_i,s)$.
In the present paper we have considered $NN$ taking both real and
imaginary parts of the optical potential as independent variables. A
much better procedure would be to constrain our fit to satisfy the
fixed-r dispersion relations. This interesting study would require
some sound assumptions on the interaction at high energies and is left
for future research.

\section{Conclusions}
\label{sec:concl}

We summarize our points. The repulsive nuclear core is a short
distance feature of the NN interaction which has been the paradigm in
Nuclear Physics for many years. This allows not only to explain
nuclear and neutron matter stability at sufficiently high densities,
but also many realistic and successful {\it ab initio} calculations
implement this repulsive core view, providing a compelling picture of
short-range correlations and even lattice QCD calculations seem to
provide evidence on the repulsive core.

In this work we have tried to verify the paradigm by analyzing
directly scattering data in a model independent way up to energies
corresponding to a wavelength short enough to separate the well
established core region from the rest of the potential. We have also
critically analyzed recent evidence provided by lattice QCD
calculations of NN potentials and seemingly supporting the repulsive
core scenario. On the light of the avoiding crossing phenomenon
triggered by multipion creation and disregarded in the lattice
calculations the repulsive core is just un upper bound of the true
energy, and not a genuine feature of the NN interaction.

A coarse grained optical potential approach has been invoked
implementing a Wilsonian re-normalization point of view. This is based
in sampling the interaction with a complex potential at points
separated by a sampling resolution corresponding to the minimal de
Broglie wavelength, $\Delta r \sim 1/p_{\rm CM}$, for a given
CM-momentum, $p_{\rm CM}$. The number of sampling points is determined
from the range of the region where the interaction is unknown. We
assume, in agreement with previous low energy studies scanning the
full database of np+pp about pion production threshold, that the
interaction above $r_c=3$fm is given by the One-Pion-Exchange
potential and sample the interaction at equidistant points separated by
the resolution scale $\Delta r$.

Traditionally, the core region is estimated to be around $a_c \sim
0.5$ fm. This requires considering LAB energies up to about 3 GeV, for
which there exist comprehensive partial wave analyses. Since, the
$^1S_0$ partial wave explores the shortest impact parameter $b \sim
1/(2p_{\rm CM})$, we have mainly restricted our analysis to this
channel and have shown that the inelasticity is concentrated in a
region below the resolution scale $\Delta r$. This amounts to treat
the inelasticity interaction as point-like and structure-less.  In the
coarse grained setup, that means that we treat only the potential at
the innermost sampling distance as complex and energy dependent. The
remaining sampling points are kept real and energy independent. We
have shown successful fits describing the real and imaginary parts of
the $^1S_0$ phase-shift up to 3 GeV LAB energy. However, two
conflicting scenarios emerge from these fits where the interaction at
short distances becomes either strongly repulsive (the repulsive core)
or strongly attractive (the structural core). Both scenarios have been
considered in the past as alternative pictures of the NN interaction
at short distances and here we show explicitly how both arise from a
similar way of sampling the interaction with a coarse grained
potential. We corroborate the main difference between both scenarios
concerning the zero energy wave function which vanishes below the core
distance in the repulsive core case and has a node at about the core
location but does not vanish below it in the structural core
situation. Since these two scenarios are phase-equivalent, we have
analyzed in a comparative way some of their features addressing the
obvious question on which one is more realistic.

There is a remarkable and tangible difference between the short
distance repulsive-core and the structural-core core scenarios,
namely, the adiabaticity of the inelasticity. In the repulsive core
case we get a rapid variation between a very strong short distance
repulsion and a very strong absorption. In contrast, in the structural
core situation, the behavior is completely smooth, and the
inelasticity couples in an adiabatic way as the energy is steadily
increased.

On the other hand, while the repulsive core has no bound states, in
agreement with experiment, the structural core presents a deeply bound
state which might pose a problem for nuclear structure calculations.
Cluster model studies have suggested that this deeply bound state is
forbidden by the Pauli principle for the underlying quarks and the
node in the wave function is just a manifestation of the composite
character of the nucleons. In fact, our way of solving the problem
using an effective interaction at the hadronic level corresponds at
the sub-nucleon level to consider just the direct Hartree term,
whereas the exchange Fock term generates naturally the node of the
relative wave function without the spurious bound state. The repulsive
core branch presents a dramatic change from the core to a strongly
absorbing potential for which we are not aware of any microscopic
explanation.

An interesting application is the study of nuclear processes where two
particles are emitted at high relative momentum and the implications
for our understanding of short distance correlations since most
studies in Nuclear Physics ignore the role of inelasticities when
fixing the interaction at high energies. The fruitfulness of the
coarse graining idea has been demonstrated recently in an efficient
solution method of the Bethe-Goldstone equation and the presently
obtained wave functions could be used to study the effects of
inelasticity on short-range correlations, a task that has never been
addressed, and is left for future research.

Finally, the present study shows that the coarse graining idea works
as expected at LAB energies as high as 3 GeV where the consideration
of inelasticities is unavoidable, and suggest to extend the present
calculation to higher partial waves or other hadronic processes.

\vskip1cm 

One of us (E.R.A.) thanks David R. Entem for discussions on quark
cluster models. We thank Jacobo Ruiz de Elvira, Quique Amaro and
Eulogio Oset for discussions and for critically reading the ms.

This work is supported by the Spanish Ministerio de Econom\'{\i}a y
Competitividad and European FEDER funds under contract
FIS2014-59386-P, FIS2014-51948-C2-1-P and by Junta Andaluc{\'{\i}a}
(grant FQM225).  P. F.-S. acknowledges financial support from the
“Ayudas para contratos predoctorales para la formación de doctores”
program (BES-2015-072049) from the Spanish MINECO and ESF.


\end{document}